# Electron diffraction data on structural transformations in free clucters of argon

Spartak I. Kovalenko, Oleksandr G. Danylchenko*, Vladimir N. Samovarov

*B. Verkin Institute for Low Temperature Physics and Engineering of the National Academy of Sciences of Ukraine,  
47 Lenin Ave., Kharkov 61103, Ukraine*

(February 10, 2004)

**Abstract**

An electron diffraction technique is used to study the structure of clusters formed in an isentropically expanding supersonic argon jet. The formation of the hcp phase with increasing cluster size is reliably detected for the first time. Observations are made for mean cluster sizes $N$ in the range from $1\times10^3$ to $8\times10^4$ atoms/cluster. An analysis of the shape of the diffraction peaks is carried out. It is found that in the range $N \leq 2\times10^3$ atoms/cluster, where the clusters are icosahedral, the profiles of the diffraction peaks are well approximated by a Lorentzian. For fcc clusters with $N \geq 3\times10^3$ atoms/cluster a better approximation is the standard Gaussian function. In the case $N \geq 1\times10^4$ atoms/cluster one observes peaks of the hcp phase in addition to the fcc peaks. The intensity of the hcp peaks increases with increasing cluster size, and for $N \approx 8\times10^4$ atoms/cluster, the (110), (101), (103), and (202) peaks, characteristic only for the hcp phase, are clearly registered in addition to the fcc peaks. A possible mechanism for the formation of the hcp structure in Ar clusters is proposed.

The dilemma of hcp-fcc structure in solidified rare gases has long been arisen but it remains still unsolved. The X-ray diffraction studies show that the rare gases (except helium) crystallize in a fcc lattice, but the calculations in the approximation of paired interaction predict the hcp structure stability. The experiments on recrystallization of the films of inert gases condensed on substrates provided some proofs for the existence of the hcp phase [1]. Evidence for a disordered hcp structure of Ar and Kr frozen in the pores of Vycor glass was obtained in [2]. For free clusters, however, the experiments on observation of the hcp phase were not performed.

The paper concerns the electron diffraction study into the structure of clusters formed within the isoentropic expanding supersonic jet of argon with Mach numbers up to 8. The experiments were carried out on a unit consisting of a generator of supersonic cluster beam and an electron diffraction device of electron energy up to 100 keV. The pumping out of the jet gas was realized with a condensation pump cooled by liquid hydrogen. The average size of clusters, $N$ (atom/cluster), was prescribed by gas pressure, $P_0$, at its constant temperature $T_0$ at the nozzle inlet. To estimate $N$, the Scherrer relation and the dependence $N=kP_0^{1.83}$, where $k$ is the constant, were used [3, 4]. The average size of clusters studied was between $1\times10^3$ and $8\times10^4$ atom/cluster. The cluster temperature in the diffraction region was about 38±4 K. The numerical values of the diffracted intensity were processed not only to construct the diffraction patterns but also to separate the background component, which is due to incoherent electron scattering and gas scattering. The diffraction pattern record was, as a rule, limited by the values of the diffraction vector $s=6$ Å$^{-1}$ ($s=4\pi sin\upsilon/\lambda$, where $\upsilon$ is the Bragg angle and $\lambda$ is the electron wave length).

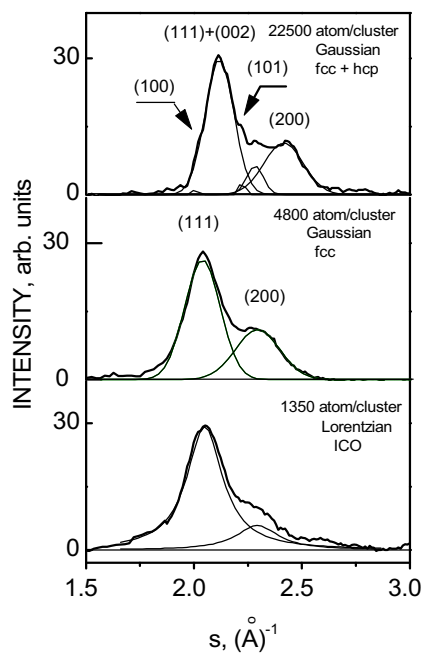

Fig. 1.

* Corresponding author. Tel.: +38-057-308576; fax: +38-057-322370; e-mail: danylchenko@ilt.kharkov.ua

The determination of the hcp structure was preceded by the analysis of diffraction peak shapes first performed for clusters of inert gases. The analysis was made for the cluster beams with $N = (1.35; 4.8; 22.5)\times 10^3$ atom/cluster. It is found that in case of $N \leq 2\times 10^3$ atom/cluster, where the clusters are of an icosahedral structure (ICO) [5, 6] the profiles of the diffraction peaks are well described by the Lorentzian contour. It turned out that for the fcc clusters ($N \geq 3\times 10^3$ atom/cluster, peaks (111) and (200)) the best approximation was the Gaussian (see Fig. 1).

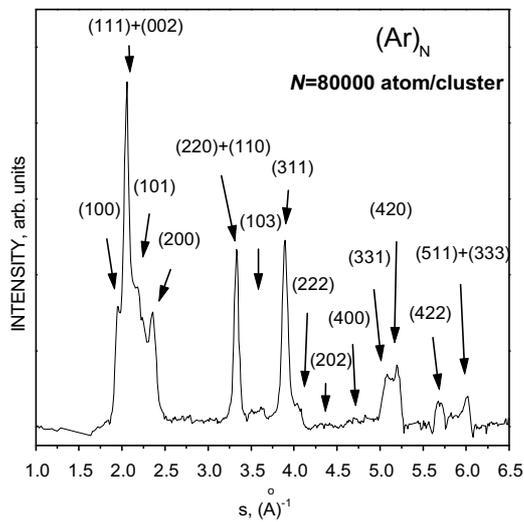

Fig. 2.

Qualitative modifications of the diffraction peak profiles are caused by the occurrence of microstresses in icosahedral clusters which are inherent of the fifth-fold symmetry structures. The microstresses produce modulations in the interatomic distances which vary from the cluster center to its surface. The modulations are responsible for the transitions from the Gaussian profile typical of clusters with constant interatomic distances to the Lorentzian profile.

We succeeded in observing the hcp phase for clusters with $N$ above $2\times 10^4$ atom/cluster. The argon clusters with $N \approx 2.2\times 10^4$ atom/cluster displayed, along with the intense fcc peaks, weak anomalies in the positions corresponding to reflexes (100) and (101) of the hcp phase (see Fig.1). The intensity of the hcp peak and, hence, the phase share contribution increased with the average cluster size. As shown in Fig. 2, the diffraction pattern for $N \approx 8\times 10^4$ atom/cluster involves well recorded hcp peaks (100), (101), (103) in addition to a number of the fcc phase ones, and with scaling the diffraction pattern up, it exhibits a slight anomaly in the intensity distribution in the region of weak hcp peak (202). A small half-width of the hcp peaks indicates that the hcp phase regions are of high linear dimensions. The increase of the hexagonal phase share contribution with average cluster size suggests that the hcp phase is formed in liquid drops and its realization is not associated with a small size of the solidified cluster.

The more detailed discussion of the obtained results will be published in Fiz. Nizk. Temp., 30 (2004) [Low Temp. Phys. 30, (2004)].